\documentclass[%
 preprint,
 amsmath,amssymb,
 aip,
floatfix,
]{revtex4-1}

\usepackage{graphicx}
\usepackage{dcolumn}
\usepackage{bm}

\usepackage{comment}
\usepackage{svg}
\usepackage{appendix}
\usepackage{amsmath}
\usepackage{amssymb}
\usepackage{xparse}
\usepackage{physics}
\usepackage{times}
\usepackage{graphicx}
\usepackage{array}
\usepackage[normalem]{ulem}
\usepackage{cmbright}

\usepackage[utf8]{inputenc}
\usepackage[T1]{fontenc}
\usepackage{mathptmx}
\usepackage{etoolbox}
\makeatletter
\def\@email#1#2{%
 \endgroup
 \patchcmd{\titleblock@produce}
  {\frontmatter@RRAPformat}
  {\frontmatter@RRAPformat{\produce@RRAP{*#1\href{mailto:#2}{#2}}}\frontmatter@RRAPformat}
  {}{}
}%
\makeatother
\begin{document}


\title{Demonstration of a droplet electrohydrodynamic blower in aerosols}
\author{Pramodt Srinivausla}
	\email{pramodt@iitb.ac.in}
	\affiliation{Indian Institute of Technology Bombay}

\author{Debabrat Biswal}
	\affiliation{Indian Institute of Technology Delhi}

\date{\today}

\begin{abstract}
Despite a rigorous analysis of nonlinear electrohydrodynamics of a liquid droplet deformation in air, its influence on the surrounding gas phase has received less attention. We observed, air circulations created due to large deformation oscillations of a pendant water droplet subject to a resonating electric field act like a micro or mini blower in the air around the droplet. This is demonstrated through the deterioration of a strong electrostatic airborne particulate capture onto the droplet, when it is placed in an aerosol. Such strong local aerodynamic influence on the surrounding aerosol particulate, heat and mass transfer in the two phase system provoke further interest in fundamental and technological research.
\end{abstract}

\maketitle 
\section{Introduction and motivation}
Attributing to the low density and viscosity of gasses as compared to liquids, momentum transfer between the two phases is often ignored on the liquid-gas interface while studying fluid dynamics of the liquid mass\cite{acheson1991elementary}. Even though the influence of gas flow on its interface with liquid is recognized in occasions such as surface waves on a liquid pool\cite{acheson1991elementary,miles1962generation} and vortices in a falling liquid droplet\cite{acheson1991elementary,hill1894vi,chapman1967formation}, dynamics in the gas phase induced by the liquid phase had received lower attention. It is the case with the decades-old extensive studies of understanding droplet electrohydrodynamic (EHD) deformation oscillations, as well\cite{sample1970quiescent,tsamopoulos1984resonant,feng1991three,depaoli1995hysteresis,trinh1996dynamics}.

Electrical Maxwell stress on the interface of two dissimilar fluids subject to an external electric field\cite{saville1997electrohydrodynamics} deforms a spherical droplet to a different deformed equilibrium capillary shape \cite{taylor1966studies,dubash2007behaviour}. The interplay of inertia of the liquid mass and capillary restoration forces on the interface during the transience of such electrostatic deformation leads to oscillations of the drop deformations. These oscillations damp out shortly due to finite viscous losses in the droplet phase\cite{ferrera2013dynamical,raisin2011electrically}. Whereas, in the case of a time-periodic AC electric field, it results in a dynamic steady state of electro-capillary deformation oscillations. Drop deformation during these oscillations amplify excessively when the harmonic electrical forcing resonates with the forced oscillations of the droplet\cite{tsamopoulos1984resonant,kang1993dynamics,trinh1994nonlinear}. Nonlinear dynamics of such large forced deformation oscillations of acoustically levitated\cite{trinh1982large}, electrostatically levitated\cite{singh2018surface}, pendant\cite{depaoli1995hysteresis,ferrera2013dynamical} and microgravity levitated\cite{trinh1994nonlinear} droplets were studied experimentally. Theoretical methods such as spherical Legendre modes decomposition \cite{trinh1982large,tsamopoulos1983nonlinear,tsamopoulos1984resonant,feng1991three} and spheroidal approximation of deformation\cite{kang1993dynamics} were employed to describe the nonlinear dynamics of drop deformation. Complete direct numerical simulations using boundary element method \cite{feng1997instability,singh2018surface}, explicit interface tracking formalism \cite{hua2008numerical} and
moving mesh method\cite{supeene2008deformation,raisin2011electrically} were reported for small deformations cases.

At large free or forced deformation of droplets, electrostatic and capillary nonlinearities manifest in a wide variety of phenomena, which were studied extensively over the last few decades. Phase plane plot of dynamics of the system, indicating variation of deformation velocity with deformation amplitude, deviate from a circular shape of linearity to elliptical shape\cite{kang1993dynamics,feng1997instability,tsamopoulos1984resonant}. Actual frequency of oscillations\cite{becker1991experimental,trinh1982large,kang1993dynamics} and hence the resonating AC frequency $f_r$\cite{feng1991three,trinh1994nonlinear} deviate from its natural capillary frequency $f_c$, depending on the electric field provoked drop deformation amplitude. Hysteresis of drop shape deformation during the forward and backward strokes in each cycle of oscillation were reported at such large deformations\cite{depaoli1995hysteresis,trinh1994nonlinear}. Hysteresis was also observed in the variation of droplet deformation response magnitude with both the frequency and magnitude of the applied electric field\cite{depaoli1995hysteresis,trinh1994nonlinear}. Beyond a limit of deformation, droplet breaks up, which is mathematically represented as a \textit{saddle bifurcation} of droplet response to the applied electric field\cite{feng1997instability}. Such disintegration modes of drop jetting or breakup at relatively lower electric stress due to the resonant inertial effects\cite{trinh1994nonlinear,feng1997instability} were studied in the interest of electrospray applications. Internal circulations in the liquid droplet were reported \cite{trinh1982large} and significant energetic coupling between different spherical modes of deformation due to such circulations were predicted at large nonlinear limits\cite{shiryaeva2003internal}.
\begin{table*}
\caption{\label{tab:DropletEHD}Typical scaling for electrohydrodynamic deformation oscillations of a water droplet in air .}
\begin{ruledtabular}
\begin{tabular}{lccc}
 & Description & Expression & Range\footnote{for drop diameter $D$ varying from $0.1mm$ to $2mm$; Distance between electrodes of potential difference $V_0=800V$ is $H=6mm$. Hence, electric field strength $E_0=1.3kV/cm$}\\
\hline
$t_c=1/f_c$ & Capillary deformation time scale & $\sqrt{\rho_l D^3/\gamma}\sim D^{1.5}$& $0.1-10 ms$ \\
$Ca_e$ & Electric capillary number & $\epsilon_g \epsilon_0 E_0^2 D/ \gamma \sim D^1$& $0.02 - 0.8$\\
   $u_{r0}$ & Velocity scale of drop tip in AC resonance oscillations & $D/t_c \sim D^{-0.5}$ & $0.85-0.19 m/s$\\
   $a_{r0}$ & Acceleration scale of drop tip in AC resonance oscillations & $D/t_c^2 \sim D^{-2}$ & $7200-18 m/s^2 $ \\
\end{tabular}
\end{ruledtabular}
\end{table*}

Consider a pendant droplet of diameter $D$ exposed to an electric field $\textbf{E}$ between a holding needle and an opposing flat plate electrode maintained at an electrical potential difference of $V_0$. Interfacial forces dominate the gravitational force for Bond number $B0= \frac{\rho_l g D^2}{\gamma}<1$, for drop diameter less than a few millimeters. Electric capillary number ($Ca_e= \frac{\epsilon_0 \epsilon_g E_0^2 D}{\gamma}$), the ratio of electric force to interfacial tension and Ohnesorge number ($Oh = \frac{\mu_l}{\sqrt{\rho_l D \gamma}}$) representing the relative influence of viscous damping over the inertial-capillary oscillations represents the operating conditions of the electro-deformation oscillations of the droplet. Here, $E_0=V_0/H$ is the nominal electric field applied and $\gamma$ is the interfacial tension. Diameter ($D$) and natural capillary frequency ($f_c$) of the droplet are used as suitable scaling for drop tip displacement and oscillation time scales respectively. Typical values of drop tip velocity and acceleration scaling are indicated in the table \ref{tab:DropletEHD} and verified from the experimental results in the later sections of this report.

With such a detailed fundamental understanding of the droplet electrohydrodynamic deformation oscillations at our disposal, we now ask, "what happens in the gas phase during this phenomenon and what can be its implications, such as on airborne particulates near the droplet ?".  

\section{Inertial oscillations of droplet electro-capillary deformation}\label{sec:dropEHDExp}
A blunt tip stainless steel needle electrode of 22 gauge size (outer diameter=0.718mm) is held vertically at a shortest distance of $6mm$ above a flat horizontal copper grounded electrode plate. Distilled water of conductivity $10^{-4}$S/m is pumped into the needle using a syringe pump to create a pendant droplet of $1.24$mm equivalent perfect spherical diameter between the two electrodes. DC or AC electric potential signal created with a function generator is amplified using a high voltage amplifier and applied between the grounded bottom plate electrode and the live needle electrode holding the droplet. Resulting droplet deformation and oscillations are observed through a microscope and captured using a high speed camera. Open source software ImageJ is used to analyze the images for the drop shape and to track the tip movement.
 
 \subsection{Case-1: constant applied electric field}
Transient deformation oscillations of the droplet observed under a constant potential difference of $800V$ between the electrodes is shown in the figure \ref{fig:DC,CFD}A, with the tip displacement and time axes scaled with drop diameter $D$ and capillary deformation time scale $t_{c}=1/f_c$ respectively.  Capillary oscillations of the droplet deformation superimposed with the transience of electrostatic stretching leads to, asymmetry of the drop tip displacement in forward and backward directions in each cycle of oscillation. 

One-fluid method using phase field equation for explicit tracking of diffused interface between two fluids is employed, to simulate the pendant drop deformation oscillations in a chamber with no slip condition on the walls. Charge conservation obeying electrostatic Poisson equation are solved with appropriate constant applied electric potential boundary conditions on the needle and plate electrode surfaces. Hydrodynamics and electrostatics are two-way coupled through electric Maxwell stresses in the Navier-Stokes equations and fluid phase fraction dependent electrical properties estimation in the electrostatic Poisson equation, as elaborated in the appendix A.
This formulation is solved numerically using finite element methods package COMSOL Multiphysics, AB/COMSOL, Inc.  Spatial variation of velocity at a time step during the forward stroke of drop oscillation, marked with a $\times$ on figure \ref{fig:DC,CFD}A) is shown on left half of the figure \ref{fig:DC,CFD}B. Magnitude of air velocity, on the left half of the figure, is observed to scale with the droplet tip velocity during the oscillations. Flow lines show how air is circulated from the drop tip to bulk air domain to the drop equator during the forward stroke. This circulation direction reverses during the backward stroke. Electric field lines, on the right half of the figure, indicate Coulombic force direction and relative magnitude distribution in the domain. 
\begin{figure}
    \centering
    \includegraphics[width=\textwidth]{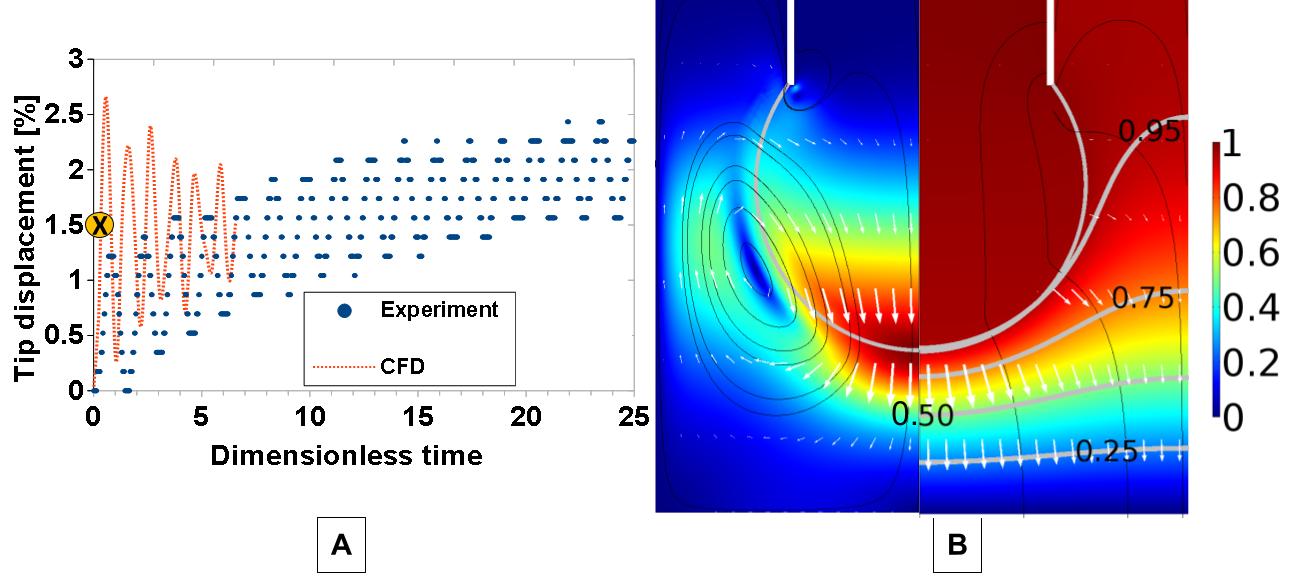}
    \caption[Pendant drop DC deformation oscillations]{Pendant drop deformation oscillations in a constant DC electric field. (A). Time series of drop tip displacement and time axes scaled with drop diameter $D$ and corresponding capillary deformation times scales $t_c$ respectively, for experiment and CFD simulation of $(Ca_e,Oh) = (0.09,0.003)$ and $(1.4,0.005)$ respectively.  (B). CFD results at scaled time t=0.25 (marked with $\times$ in figure A)- In the left half, color, arrows and curves indicate normalized velocity magnitude, direction and streamlines of flow, respectively. In the right half, color, arrows and black curves indicate normalized magnitude, local directions and field lines of electric field respectively. Contours of equi-electric potential are labeled with corresponding values of $V/V_0$.}
    \label{fig:DC,CFD}
\end{figure}
\subsection{Case-2: AC electric field resonating with droplet deformation oscillations\label{subsec:ACosc}}
\begin{figure*}[!htbp]
\includegraphics[width=1\textwidth]{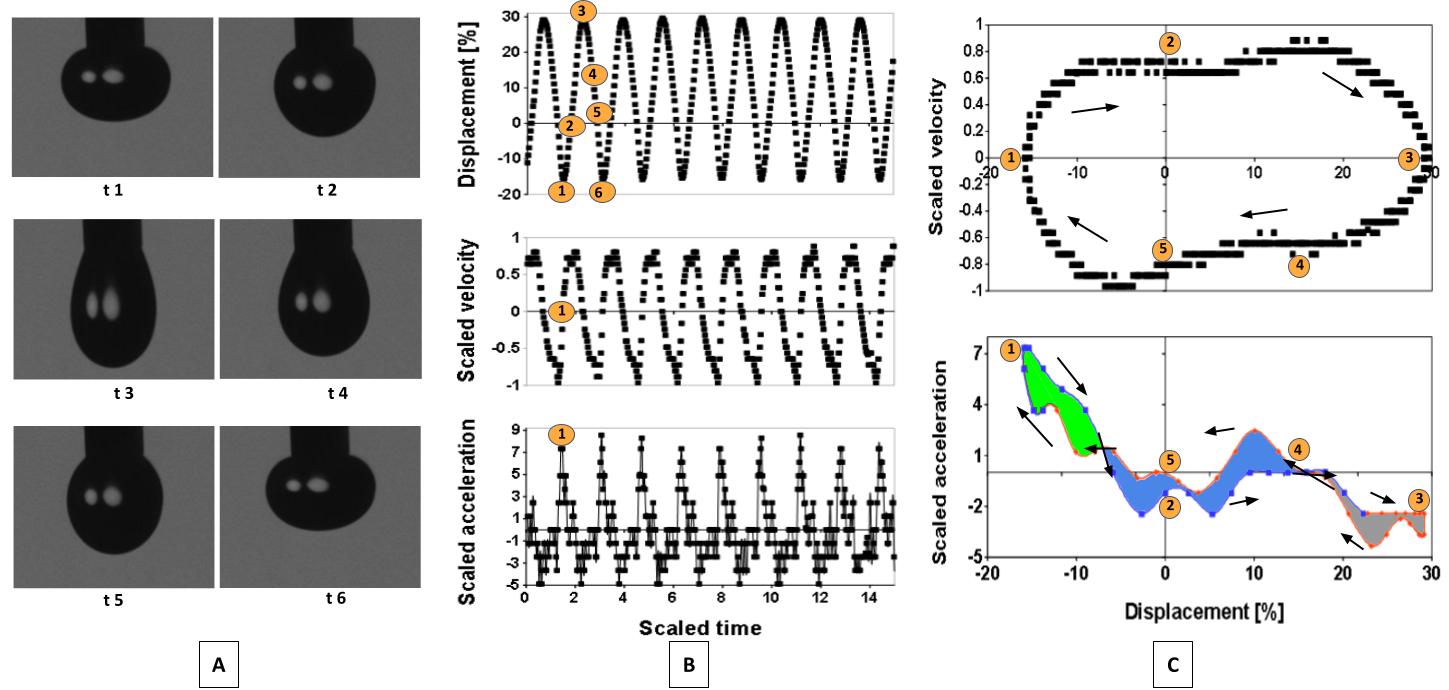}
\caption[Pendant drop AC resonance deformation oscillations]{\label{fig2:AC} Drop deformation oscillations during harmonic steady state under a resonating AC forcing. (A). Time lapse images of drop deformation (B). Time series of dimensionless drop tip displacement ($\hat{x}_{r(t)}$), velocity ($\hat{u}_{r(t)}$) and acceleration ($\hat{a}_{r(t)}$) scaled with the capillary deformation scaling indicated in table \ref{tab:DropletEHD}. (C). Phase plane plots of drop tip velocity ($\hat{u}_{r(t)}$) and acceleration ($\hat{a}_{r(t)}$) variation with drop tip displacement (in one cycle of deformation marked in B) }.
\end{figure*}

An AC signal of  800V rms potential difference was applied, on the same sized droplet under the same conditions of case 1, at corresponding resonance frequency, identified as $f_{AC}=63Hz$. Resulting dynamic steady state of drop deformation oscillations with tip displacement more than 10 times that of DC case are presented in figure \ref{fig2:AC}. Drop oscillations are observed to be at double the applied AC frequency and higher than the drop natural capillary frequency, $f_r = 2 f_{AC}=1.35 f_c$. Figure \ref{fig2:AC}A indicates the large deformations of droplet from oblate to prolate and back to the oblate shapes during the forward and backward strokes respectively, at time steps as marked on figure \ref{fig2:AC}B. Hysteresis of shape deformation between forward stroke, at time steps $t2$ and backward stroke $t5$ can be noticed, with no significant drop volume variation throughout the axisymmetric shape oscillations. In spite of the large deformation, shape deformation observed is dominantly similar to the P2 mode of spherical deformation of radial coordinate $r$ of the surface with the angular coordinate $\theta$, $r_{(\theta,t)}= r_0(1+\alpha^{(2)}_{(t)} \frac{(3 cos^2(\theta)-1)}{2})$; where $\alpha_{(t)}$ is the time dependent deformation. Surface waves consisting of significant contribution from higher modes were observed only at much higher applied AC frequency, which is far away from the resonance window. 

Time series of drop tip displacement, velocity and acceleration non-dimensionalized with corresponding scaling indicated in table \ref{tab:DropletEHD}, are shown in figure \ref{fig2:AC}B. This confirms the oscillations to be at the capillary time scale $t_c$ and at length scale of the diameter of the droplet $D$. However, the drop tip has higher displacement in the forward stroke than the backward stroke, due to electrostatic force in the constant stretching direction. Although velocity amplitudes in both forward ($+$ve) and backward ($-$ve) directions are nearly the same, broken symmetry of their variation during the two strokes can be noticed. This results in a significant difference in the tip acceleration in the forward and backward directions. In the prolate shape position of the drop, interfacial tension and electrostatic forces oppose each other, while they are collaborative in the forward($+$ve) direction creating a recoil of highest acceleration. Elliptical shape of the velocity phase plane, shown in figure \ref{fig2:AC}C indicates the deviation from linear harmonic oscillations due to large deformation. Acceleration phase plane plot of data from any arbitrarily selected cycle of oscillation indicates hysteresis in the three phases of the oscillation. Net positive work done on the droplet during prolate ($+$ve) and oblate ($-$ve) shape phases are nearly counterbalanced by the negative work done during the neutral shape phase. The net work done is dissipated as viscous losses, dominantly in the liquid phase circulations and also as a net forward impact on the surrounding air.

\subsection{Air blowing effect of droplet electrohydrodynamics\label{subsec:ACblower}}
Newtonian formulation estimates drag on a rigid curved surface by surrounding gas, as its inertial momentum exchange with an equivalent projected area of the surface\cite{hinds1999aerosol}. Similarly, force exerted by the moving droplet on air (of density $\rho_g$) while displacing it at up to a rate of, $\dot{m}_0= \rho_g \frac{\pi}{4} D^2 v_0$ is $F_{d0}= C_D \rho_g \frac{\pi D^2}{4} v_0^2$\cite{hinds1999aerosol}. Hence, pressure on air is of the order of inertial pressure scaling $P_0= \rho_g v_0^2$. Force imparted on air in each stroke of droplet of equivalent length of $x_s$ is  $F_{\pm}=m_g a_{\pm}$; where $m_g = \rho_g x_s \pi D^2/4$, is the mass of gas displaced in each stroke, $x_s=\eta_s x$ is the equivalent displacement of projected area and $a_{\pm}$ is the acceleration of the drop projected area in the forward (+) or backward (-) direction. From the experimental results, shown in figure \ref{fig2:AC}, nonlinearity of large deformation oscillations leads to asymmetry in the acceleration between forward and backward strokes, $\Delta a=(a_+-a_-)\sim a_0$, while displacement of the forward and backward strokes are equal. Hence a \textit{differential blow pressure}, $P_B \sim\frac{1}{2} \rho_g \hat{x}_{r(Ca_e)} D \Delta a$ is generated, which effectively scales as $D^{-1}$. Ratio of this pressure differential with capillary pressure scale, $\hat{P_B}= P_B/(\gamma/D)= (\rho_g/\rho_l) \frac{ \hat{x}_r}{2}$ is dependent only on the scaled stroke length of electrocapillary oscillations of the droplet. 

\section{Droplet oscillations in an aerosol chamber}
\begin{figure}[!htbp]
\includegraphics[width=\textwidth]{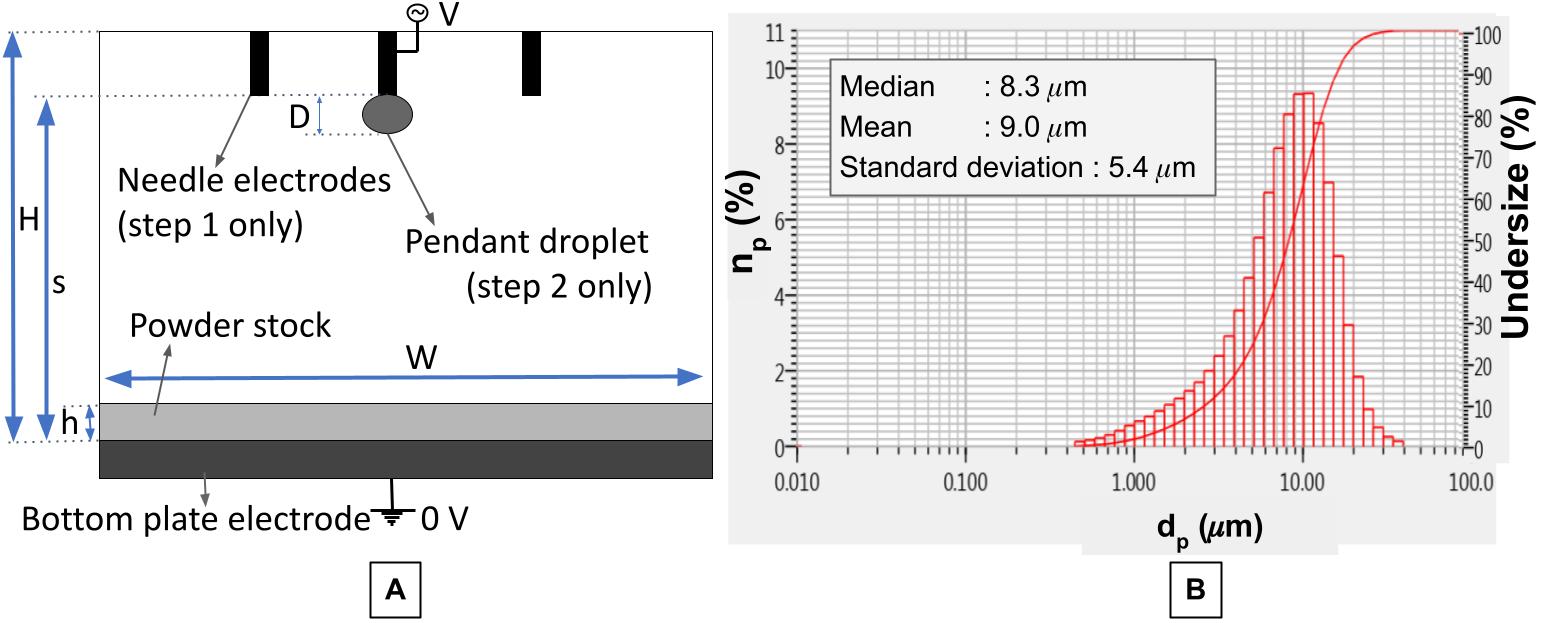}
\caption[Experimental setup of aerosol chamber]{\label{fig1:setup} Experimental setup of aerosol chamber.  (A). Closed aerosol chamber is of dimensions W=H=$10mm$ and chamber length perpendicular to the plane of the image L=$15mm$, with transparent side walls, bottom metal plate electrode and top face holding needle electrodes. A layer of particulate powder is spread on the bottom plate to a height  h=$1mm$. In step 1 of the experiment, three metal needles are placed from top phase to fluidize particulates into aerosol and then removed. In step 2 of the experiment, one fresh metal needle electrode is inserted and the pendant droplet is created to a shortest distance from the bottom plate s=$6mm$, as shown. (B). Particle size distribution with wet laser scattering technique using Horiba LA-960. Median particle size is $8.3\mu m$ and standard deviation is $5.4\mu m$ with $90\%$ particles in the range of $2-20\mu m$.
}
\end{figure}
Following experiment is conducted to study the effect of aerodynamic circulations created by the droplet EHD oscillations, on the aerosol particulate near the droplet.
Setup contains a closed chamber with copper grounded electrode plate at the bottom and transparent acrylic walls on all sides. One of the side walls is replaced with a transparent removable tape to allow physical access in to the chamber. Top face of the chamber holds needle electrodes vertically down into the chamber. 
Incense sticks are burned separately, to release sub micron particles as smoke. Remaining bottom ash is dried and sieved to collect particles of Gaussian size distribution shown in figure 1B with median size around $8.3\mu m$ measured using wet laser scattering technique of \textit{Horiba LA-960} aerosol sizer equipment. $50mg$ of these hydrophilic particles sample is spread uniformly on the clean bottom electrode plate, in each run of the experiment in the clean air chamber. Major constituents of incense stick are bamboo stick (density~$~0.8g/cc$), charcoal ($~1-2.5g/cc$) and saw-dust wood powder. Hence density of particles is considered to be about$~1.2g/cc$ \cite{ji2010characterization}.

\subsection{Step-1: Aerosol generation in the observation chamber, using electrostatic technique}
Creating aerosol often requires specialized and sophisticated equipment. However, an easy and compact technique to create aerosol directly in the observation chamber utilizing the same electrical setup of the experiment is exercised. Electric field is often used in powder technologies to momentarily liftoff particulate matter larger than airborne size, to achieve uniform spreading of powder on a conveyor\cite{shoyama2018particle} and to study the interaction of particles placed on a plate electrode with spherical liquid electrodes\cite{zuo2016particle}. It has been concluded that particles lift off due to Coulombic force\cite{shoyama2018particle,cho1964contact} in such settings. Dielectrophoretic (DEP) force is weak and electrostatic image forces are significant only when the particles are close to the electrodes. In these techniques, particles placed on a bottom plate electrode attain charge from the bottom electrode plate, which then lift them off to fly towards the opposite polarity needle or drop electrode. Significant fraction of the charge attained by submicron particles transfers from the electrode through contact charge mechanism. On the other hand, charging by induction is known to dominate the contact charging for micron sized particles\cite{cho1964contact}. For more details on the charging mechanics, refer to suitable reviews on tribo or contact electric charging \cite{matsusaka2010triboelectric}. Cho et.al., \cite{cho1964contact} reported the specific charge increases and the duration to attain the charge decreases, with increasing conductivity of the particle.  Cho et.al., \cite{cho1964contact} also observed particle dynamics in flight decrease as conductivity decreases from conductors to semi to insulators. Zuo et.al.\cite{zuo2016particle} estimated experimentally and numerically, specific charge equivalent to thousands of electrons is attained on each particle of above micron size. 

\begin{figure}
    \centering
    \includegraphics[width=\textwidth]{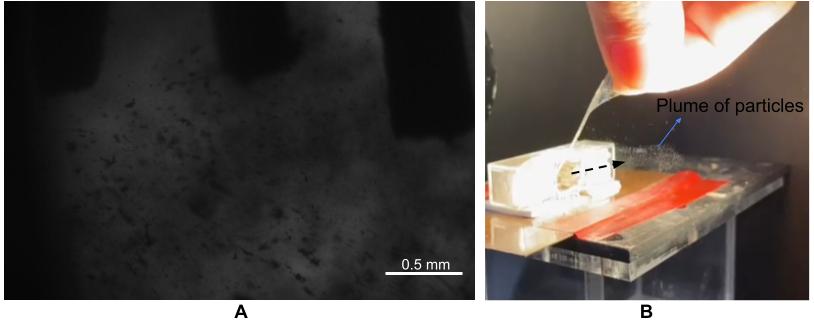}
    \caption{(A). Explosive lift off of particles from the ground plate, generating aerosol. (B). Demonstration of aerosol particles in the chamber, with a release of airborne particulate up on opening a side wall of the chamber a few minutes after turning off the electric field. Videos are presented in supplementary materials 1 \& 2 respectively.}
    \label{fig:Fluidize}
\end{figure}
A simple electrostatic technique is synthesized in a similar way, to disperse particles spread on the bottom ground plate into the experimental chamber. However  a pendant droplet would be destabilized by such a high electric field to breakup into a spray. Hence a two step process is followed. In the first step,  three metal needle electrodes equally spaced spanning along the length of the chamber are placed from the top, to a closest separation of 8mm from the bottom grounded electrode plate. An electric potential difference of $8kV$ is applied between the needles and the bottom grounded plate. Thus a nominal electric field of $10kV/cm$  is applied in the chamber, for a duration of $30s$ without creating an electric arc. This field is about 10 times stronger than that of the drop deformation oscillations case discussed in section 2 and higher than that used in literature mentioned above to lift off powder particles\cite{shoyama2018particle,zuo2016particle}).
Incense stick ash particles lift off explosively, after about 1-2 sec of particles charging delay, after applying the electric field, as indicated in the figure \ref{fig:Fluidize} with supplementary material 1, video. Same experiment using Silane coated soda lime particles takes more than 10 sec before their milder lift-off is observed, confirming a higher electric conductivity of incense stick particles. Multiple such explosions of incense stick particles take place, with increasing duration of gap between consequent explosions. 

During the particulate explosive liftoff, particle agglomerates moving at high velocities disintegrate while moving\cite{shoyama2019mechanism}and bounce-off from the electrodes due to the impact\cite{zuo2016particle} and contact charge repulsion\cite{shoyama2019mechanism}, blowing them explosively everywhere in the chamber, as seen in the figure \ref{fig:Fluidize} with supplementary material 2, video. After the electric field is turned off, charged particles which contact the side walls of the chamber stick on to them while those which are airborne in the chamber remain airborne. Upon opening the side wall closure tape 2 minutes after the electric field is turned off, particles airborne in the chamber diffuse out as shown in the figure \ref{fig:Fluidize} with corresponding supplementary material video.

\subsection{Step-2: Electrostatic capture of particles subject to droplet oscillations}
Then, the second step of the experiment is to create and electrify a pendant droplet in this aerosolized chamber, with the same droplet size, separation from bottom electrode and DC or equivalent resonating AC electrical signals as that of section \ref{sec:dropEHDExp}. In this step of the experiment, electrodes used for fluidizing the particles are immediately replaced with a single clean needle from the top face, without leaking out the aerosol in the chamber. 
Droplet of the same size and their oscillations shown in figures \ref{fig:DC,CFD} and \ref{fig2:AC} were observed at the same applied electrical operating conditions in the aerosol chamber. These electric fields are less than 1/10th of that used for fluidizing the particles and hence not anticipated to influence the particles from the bottom or side walls but only influence the airborne particles around the droplets. 

Momentum balance for particle of mass, density, relative permittivity $m_p,\rho_p,\epsilon_p$ and velocity $\textbf{u}_p$in an electric field \textbf{E} and air flow velocity field \textbf{u} is,
\begin{align}
m_p \pdv[2]{\textbf{x}_p}{t} =& \frac{m_p}{\tau_p} \frac{C_D}{24/Re_p} (\textbf{u}-\textbf{u}_p)+m_p \textbf{g}+ q_p \textbf{E} \nonumber\\
&+ \frac{\pi \epsilon_g \epsilon_0\kappa_{CM}}{4} d_p^3 \nabla \textbf{E}^2+ \frac{q_p^2}{16 \pi \epsilon_g \epsilon_0 z_p^2}
\end{align}
Where, $\tau_p = \frac{\rho_p d_p^2}{18 \mu_g}$ is the particle viscous relaxation time scale, $Re_p =\frac{\rho_p (v-v_p) d_p}{\mu_g} \leq 1000$ is the particle Reynolds number. Value of drag coefficient $C_D$ can be found from Morsi et.al\cite{morsi1972investigation} as $C_D = \frac{24}{Re_p} (1+0.15 Re_p^{0.687})$ and Clausius-Mossotti factor of particle polarization in air is $\kappa_{CM} = \frac{\epsilon_p-\epsilon_g}{\epsilon_p+2 \epsilon_g} $with particle and air relative permittivities $\epsilon_p, \epsilon_g$. Forces on the right hand side of the equation are aerodynamic drag, gravity, electrostatic Coulombic, dielectrophoretic (DEP) and image forces respectively\cite{shoyama2018particle}. Particle charge conduction time scale  $t_{\sigma}=\epsilon_0 \epsilon_p/\sigma_p\sim 1\mu s$  is too short compared to the AC field time scale $t_{AC}=1/f_{AC}=16ms$ for the time variation of electric field to affect particles, any differently than constant potential of DC. While the air circulations impart momentum to particle through drag force, their inertia and electrostatic forces deviates their path from the streamlines. Ratio of different forces indicate their relative importance. Stokes number, indicating the finite inertial effect over the drag force at the time scale of air circulations, $St_c=\frac{\tau_p}{t_c}\approx0.1$ for the particles of $8.3 micron$ size and $1200kg/m^3$ density. Ratios of inertial influence with Coulombic and DEP forces on the particles charged with $1000$ electrons are, $\Xi_{c}= \frac{\pi \rho_p d_p^3 D}{6 z_p e E_0 t_c^2}\sim \frac{1}{D^2} \approx 0.8$ and $\Xi_{u}= \frac{3}{2} \frac{\epsilon_0 \epsilon_g \kappa_{CM} E_0^2 t_c^2}{\rho_p D^2} \sim D  \approx400$ respectively. Image forces are not important, as airborne particles are farther from the metal electrodes compared to their size, during most of the time. Hence inertia and Coulombic force are the most dominant influences on the particles, the balance between which determines the fate of the airborne particles in the domain.

After a duration of $60$ seconds of electrifying the droplet in the chamber, electric field is turned off and the droplet is collected on a glass slide. It is then evaporated at a temperature of $100^0C$ in a clean closed oven and the particle deposition on the glass slide is observed through a microscope, which are the aerosol particulate collected onto and into the droplet from the aerosol chamber.
\begin{figure}[!htbp]
\includegraphics[width=0.6\textwidth]{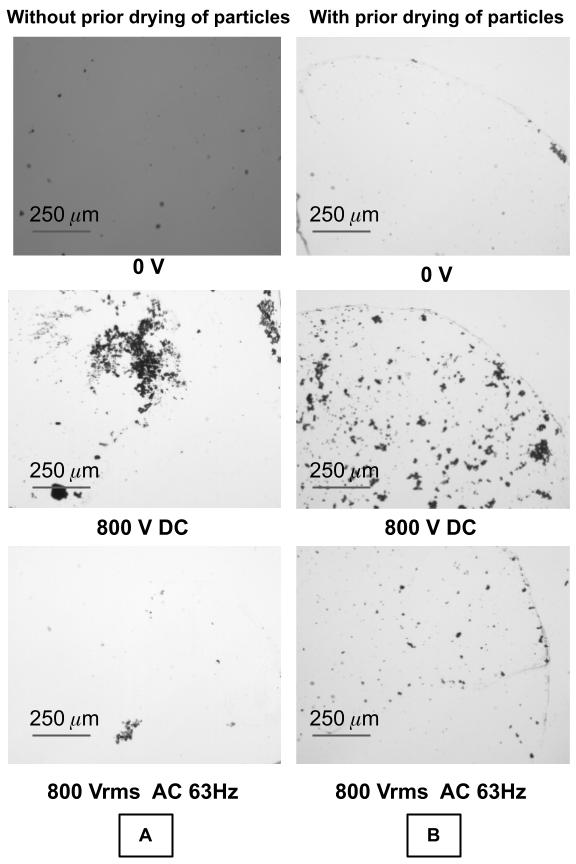}
\caption[Microscopy result of particulate collected in the droplet]{\label{fig3:microscopyresult} Microscopy result of particulate collected in the droplet subject to no electric field, DC and resonating AC electric fields when (A). particles used without and (B). after prior drying. DC case indicates more particle capture than AC and dried particles have more capture on droplet than un-dried particles.}
\end{figure}
Typical results of this experiment are shown in the figure \ref{fig3:microscopyresult}, which indicate a distinct deterioration of particle capture in case of AC field droplet oscillations, as compared to DC. AC case results are comparable to that of zero electric field in stagnant air chamber. Hence air circulations originating from the electrified droplet interface counteract electrostatic attraction of particles and act as a protective layer from aerosol particulate around it. Similar results of how relative velocity between oppositely charged particles and a falling droplet deteriorates the particulate capture were reported by Admiak et.al.\cite{adamiak2001deposition}. Also, this experiment is repeated with the same particulate sample, after a prior drying in an oven to remove moisture in them to enable easier breakup of particulate agglomerates to create aerosol in the chamber. This results in a higher particulate capture, in all the three cases of electrical signals. 


 Value of $\Xi_c\sim1$ indicates strong influence of droplet blower air circulations comparable to that of electrostatic particle attraction. Spatial and temporal variation of the velocity field result in deterioration of particle capture onto the oscillating drop electrode.
 Figure \ref{fig:DC,CFD}B indicates spatial variation of velocity field with perturbation due to small deformation oscillations ($\sim 2\%$ of AC resonant oscillations). This indicates velocity magnitude decreases with distance from the drop interface. Due to particle inertia, this distribution may result in a net outward inertial movement of particle, that counter balances the electrostatic attraction. Such a phenomenon is visualized for large dust particles suspended in oil in front of an oscillating water droplet, described in appendix\ref{Append:DustOil}. In addition to that, nonlinearity of temporal variation of drop tip movement at higher order, identified in terms of drop tip acceleration in section \ref{subsec:ACosc}, has a net effect of blowing away the particles in each cycle of oscillation; Hence keeping the particles oscillating airborne rather than collected on the drop surface.

\section{Prospective applications}
Interplay of aerodynamic circulations and electrostatic field on aerosol particulates has been recognized as an important phenomenon in a variety of applications\cite{Srinivasula2022airborne,davenport1978field}. Wet walls \cite{lin2010efficient} and membrane walls\cite{phadke2021novel,bayless2004membrane} were used as collection and transfer media of airborne particulate in inertial capture systems. Droplet EHD blowers may be useful to improve local delivery of natural uncharged or mildly charged airborne particles onto them, hence avoid air ionization and Ozone generation for particulate capture in filterless electrostatic precipitators ESPs. Electrospray of micro charged liquid droplets is also used to avoid air ionization and hence Ozone release in wet ESPs\cite{tepper2007electrospray}. Local vortices created by milli size pendant droplet oscillations may be utilized to improve the particulate - micro droplets airborne interaction. However liquid consumption is a bottleneck for this technique.  A self-cleaning air purifier design was presented by  Srinivasula et.al.\cite{Pramodt2022patent} where arrays of oscillating and stagnant anchored electrified droplets were used both as air circulation generators and particulate collectors, which are periodically recirculated and reconditioned in a closed external water circuit.
\begin{figure}
    \centering
    \includegraphics[width=0.5\textwidth]{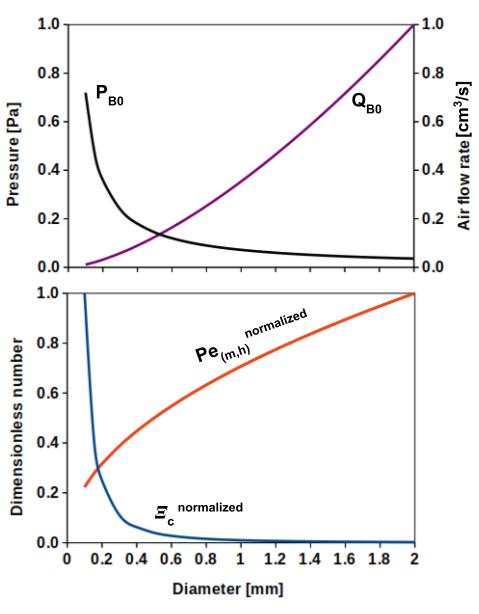}
    \caption{Drop size dependence on variation of  droplet blower characteristic performance, influence on the aerosol, heat and mass transfer. Variation of working pressure scaling $P_{B0}$, air flow rate scaling $Q_{B0}$ and characteristic dimensionless numbers $Pe_{(h)}, Pe_{(m)}$ and $\Xi_c$ normalized with $15, 2\times 10^5$ and $120$ respectively.}\label{fig:scalings}
\end{figure}

More applications, such as in milli/micro scale heat, species and gas transport problems may be recognized. Suitable representative dimensionless numbers scaling are shown in the figure \ref{fig:scalings}. Variation of operating pressure scale $P_{B0}= \frac{\rho_g D^2}{t_c^2}\sim D^{-1}$ and air flow rate delivery scale $Q_{B0}=\frac{\pi D^3}{4t_c} \sim D^{1.5}$ of such a pump, with the droplet size is shown in figure \ref{fig:scalings}. By incorporating suitable configuration of MEMS gas one way valves as inlet and outlet flow controllers, these oscillating droplets may be engineered into a mini reciprocating air pump.
Improving heat transfer at milli and micro length scales using noninvasive controllable electrostatic techniques is a popular area of active research\cite{go2007ionic,go2008enhancement}. Improvement in heat transfer due to electrohydrodynamic circulations in a drop during transient DC deformation was numerically reported recently by Jiang et.al. \cite{jiang2022numerical}. Peclet number of heat transfer in air in case of AC resonant oscillations of a droplet, $Pe_{h}= \frac{\rho_g c_{pg}D^2}{k t_c} \sim 10$ represents heat transfer enhancement in air phase as well, due to local air convection. Here, $c_{pg}$ and $k$ are the specific heat capacity and thermal conductivity of air. Improvement in gaseous species transfer from air to droplet phase due to circulations in the droplet during its electrostatic deformation was reported by Carleson et.al. \cite{carleson1983effect}. A finite component of drop and bulk phase velocity tangential with their interface can be seen in the figure \ref{fig:DC,CFD}B, which improves the advection of dissolvable gas into the droplet due to water circulations. Peclet number of mass transfer in the AC resonant droplet case is $Pe_{m}=\frac{D^2}{t_c D_g}$, where $D_g$is the coefficient of diffusion of a water dissolvable gas in air. Variation of  these dimensionless numbers, Peclet numbers for heat transfer $Pe_h$, mass transfer $Pe_m$ and inertial-Coulombic ratio number $\Xi_c$normalized with $15$, $2\times10^5$ and 120 respectively, with drop size at resonant oscillations is shown in figure \ref{fig:scalings}. Air thermal diffusion is  $\alpha_g= \frac{rho_g c_{pg}}{k_g}= 2.6\times10^{-5} m^2/s$, gas diffusion in air is considered as that of $CO_2$ $D_{g}= 1.9\times 10^{-9}m^2/s$. Charge on particles of $8.3\mu m$ diameter, $1200kg/m^3$ density is considered as $1000$ electrons.

\section{Conclusion and future directions}
A new perspective to connect droplet electrohydrodynamics with airborne particulate is demonstrated in this work using experimental observation of large nonlinear deformation oscillations of a water droplet in an enclosed particulate aerosol chamber. Strength of the influence of air circulations created by the micro/milli scale oscillating droplet blower during such oscillations, on airborne particles is measured in terms of counter balancing strong Coulombic attractive force. While scope of this article is to motivate the interdisciplinary opportunity of combining droplet EHD with particulate physics \& electrostatics through demonstrative experiments, understanding the detailed and quantitative mechanics of the phenomena remains as scope for future investigations. Direct numerical methods to simulate AC resonance deformation oscillations of droplets is still a state-of-the-art challenge for the computational community. PIV methods for particle tracking visualization and a theoretical framework to combine droplet EHD with airborne particle electrostatics would be a useful guide for further research.

\section{Author declarations }
\subsection{Conflicts of interest}
The authors have no conflicts of interest to disclose.
\subsection{Authors' contributions}
PS conceived, conducted and reported the research. DB contributed to designing and implementing the experiments.
\subsection{Acknowledgements}
Authors thank Prof.Rochish Thaokar for providing his valuable suggestions and laboratory resources. PS acknowledges Department of Science and Technology (DST), India for financial support during the period of research.
\subsection{Data availability}
The data that support the findings of this study are available from the corresponding author
upon reasonable request.

\appendix
\section{CFD methodology}
The following equations were used to solve the hydrodynamics of the multiphase system problem as described in the COMSOL documentation on theory for two phase flows\cite{phasefCOMSOL}, 
\begin{align}
\nabla.\textbf{u}&=0\\
\rho_f \frac{D\textbf{u}}{Dt} &= \nabla.\left( \tau_M + \tau_E \right) + F_{ST}
\end{align}
Where, mechanical stress tensor ${\tau}_M = -p\textbf{I} + \eta_f(\nabla \textbf{u}+(\nabla \textbf{u})^T)$ and electrical stress tensor $\tau_E=\epsilon_0 \epsilon_f(\textbf{E}\textbf{E}- \frac{E^2}{2}\textbf{I})$. $\textbf{F}_{ST}$ is the volume formulation of surface tension force using a scalar variable field $\phi$, called phase field, evaluated from
\begin{align}
\frac{D\phi}{Dt} &=\frac{3}{2\sqrt{2}} \nabla. (\chi  \alpha_p \gamma \nabla \psi) \\ 
\psi &= - \nabla. \alpha_p^2 \nabla \phi + (\phi^2-1)\phi
\end{align}
$\phi$ has uniform values of $\pm 1$ in liquid droplet and bulk air respectively, with a transition layer of thickness $\alpha_p$ on the interface of surface tension $\gamma$. Value of the mobility parameter $\chi$ is adjusted to introduce an optimal value of artificial interface diffusion across the actual sharp interface, to enable smooth computation of the multiphase system. Then the surface tension force is estimated as, $\textbf{F}_{ST}= \frac{3 \alpha \gamma}{2 \sqrt{2}} \nabla \phi (-\nabla^2 \phi + \frac{\phi(\phi^2-1)}{\alpha^2})$.

Electrical charge conservation combined with the Poisson equation of electrostatics, indicated below, is solved for the electric field distribution $\textbf{E}$. 
\begin{align}
\nabla. \left( \epsilon_0 \epsilon_f \pdv{\nabla V}{t} + \sigma_f \nabla V\right) =0
\end{align}
Local mechanical and electrical properties of the fluid is estimated as an arithmetic and harmonic interpolation of the liquid and gas phases weighted with the phase volume fraction\cite{tomar2007two}.
\begin{align}
\rho_f = \rho_l \text{Vf}_l + \rho_g \text{Vf}_g \quad &; \quad \eta_f = \eta_l \text{Vf}_l + \eta_g \text{Vf}_g \\
\epsilon_f = \frac{\text{Vf}_l}{\epsilon_l} + \frac{\text{Vf}_g}{\epsilon_g} \quad&; \quad \sigma_f = \frac{\text{Vf}_l}{\sigma_l} + \frac{\text{Vf}_g}{\sigma_g}
\end{align}
\section{Suspended dust particle dynamics in oil around an oscillating droplet\label{Append:DustOil}}
\begin{figure}
    \centering
    \includegraphics[width=0.5\textwidth]{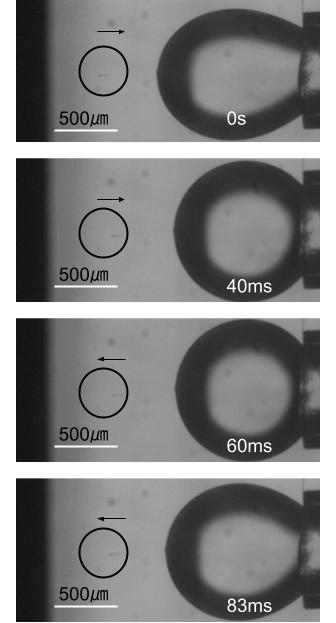}
    \caption{Suspended dust particle dynamics in 20cSt viscosity Silicone oil, around an oscillating water droplet subject to resonating AC field. Refer supplementary material for video.}\label{fig:dustAppend}
\end{figure}
Figure \ref{fig:dustAppend} indicates a dust particle suspended in 20 cSt Silicone oil in front of a horizontal water drop oscillating subject to resonant AC electric field. Refer to the video supplementary material 3 \& 4 videos for the complete recorded experimental observation.
\bibliography{aipsamp}

\begin{thebibliography}{10}

\bibitem{acheson1991elementary}
D.~J. Acheson, ``Elementary fluid dynamics,'' 1991.

\bibitem{miles1962generation}
J.~W. Miles, ``On the generation of surface waves by shear flows. part 4,''
  {\em Journal of Fluid Mechanics}, vol.~13, no.~3, pp.~433--448, 1962.

\bibitem{hill1894vi}
M.~J.~M. Hill, ``Vi. on a spherical vortex,'' {\em Philosophical Transactions
  of the Royal Society of London.(A.)}, no.~185, pp.~213--245, 1894.

\bibitem{chapman1967formation}
D.~S. Chapman and P.~Critchlow, ``Formation of vortex rings from falling
  drops,'' {\em Journal of Fluid Mechanics}, vol.~29, no.~1, pp.~177--185,
  1967.

\bibitem{sample1970quiescent}
S.~B. Sample, B.~Raghupathy, and C.~D. Hendricks, ``Quiescent distortion and
  resonant oscillations of a liquid drop in an electric field,'' {\em
  International Journal of Engineering Science}, vol.~8, no.~1, pp.~97--109,
  1970.

\bibitem{tsamopoulos1984resonant}
J.~A. Tsamopoulos and R.~A. Brown, ``Resonant oscillations of inviscid charged
  drops,'' {\em Journal of Fluid Mechanics}, vol.~147, pp.~373--395, 1984.

\bibitem{feng1991three}
J.~Q. Feng and K.~V. Beard, ``Three-dimensional oscillation characteristics of
  electrostatically deformed drops,'' {\em Journal of fluid mechanics},
  vol.~227, pp.~429--447, 1991.

\bibitem{depaoli1995hysteresis}
D.~DePaoli, J.~Feng, O.~Basaran, and T.~Scott, ``Hysteresis in forced
  oscillations of pendant drops,'' {\em Physics of Fluids}, vol.~7, no.~6,
  pp.~1181--1183, 1995.

\bibitem{trinh1996dynamics}
E.~Trinh, R.~Holt, and D.~Thiessen, ``The dynamics of ultrasonically levitated
  drops in an electric field,'' {\em Physics of Fluids}, vol.~8, no.~1,
  pp.~43--61, 1996.

\bibitem{saville1997electrohydrodynamics}
D.~Saville, ``Electrohydrodynamics: the taylor-melcher leaky dielectric
  model,'' {\em Annual review of fluid mechanics}, vol.~29, no.~1, pp.~27--64,
  1997.

\bibitem{taylor1966studies}
G.~I. Taylor, ``Studies in electrohydrodynamics. i. the circulation produced in
  a drop by an electric field,'' {\em Proceedings of the Royal Society of
  London. Series A. Mathematical and Physical Sciences}, vol.~291, no.~1425,
  pp.~159--166, 1966.

\bibitem{dubash2007behaviour}
N.~Dubash and A.~Mestel, ``Behaviour of a conducting drop in a highly viscous
  fluid subject to an electric field,'' {\em Journal of Fluid Mechanics},
  vol.~581, pp.~469--493, 2007.

\bibitem{ferrera2013dynamical}
C.~Ferrera, J.~L{\'o}pez-Herrera, M.~Herrada, J.~Montanero, and A.~Acero,
  ``Dynamical behavior of electrified pendant drops,'' {\em Physics of Fluids},
  vol.~25, no.~1, p.~012104, 2013.

\bibitem{raisin2011electrically}
J.~Raisin, J.-L. Reboud, and P.~Atten, ``Electrically induced deformations of
  water--air and water--oil interfaces in relation with electrocoalescence,''
  {\em Journal of Electrostatics}, vol.~69, no.~4, pp.~275--283, 2011.

\bibitem{kang1993dynamics}
I.~Kang, ``Dynamics of a conducting drop in a time-periodic electric field,''
  {\em Journal of Fluid Mechanics}, vol.~257, pp.~229--264, 1993.

\bibitem{trinh1994nonlinear}
E.~H. Trinh, L.~Leal, Z.~Feng, and R.~Holt, ``Nonlinear dynamics of drops and
  bubbles and chaotic phenomena,'' in {\em NASA. Lewis Research Center, Second
  Microgravity Fluid Physics Conference}, 1994.

\bibitem{trinh1982large}
E.~Trinh and T.~Wang, ``Large-amplitude free and driven drop-shape
  oscillations: experimental observations,'' {\em Journal of Fluid Mechanics},
  vol.~122, pp.~315--338, 1982.

\bibitem{singh2018surface}
M.~Singh, N.~Gawande, Y.~Mayya, and R.~Thaokar, ``Surface oscillations of a
  sub-rayleigh charged drop levitated in a quadrupole trap,'' {\em Physics of
  Fluids}, vol.~30, no.~12, p.~122105, 2018.

\bibitem{tsamopoulos1983nonlinear}
J.~A. Tsamopoulos and R.~A. Brown, ``Nonlinear oscillations of inviscid drops
  and bubbles,'' {\em Journal of Fluid Mechanics}, vol.~127, pp.~519--537,
  1983.

\bibitem{feng1997instability}
Z.~Feng, ``Instability caused by the coupling between non-resonant shape
  oscillation modes of a charged conducting drop,'' {\em Journal of Fluid
  Mechanics}, vol.~333, pp.~1--21, 1997.

\bibitem{hua2008numerical}
J.~Hua, L.~K. Lim, and C.-H. Wang, ``Numerical simulation of deformation/motion
  of a drop suspended in viscous liquids under influence of steady electric
  fields,'' {\em Physics of Fluids}, vol.~20, no.~11, p.~113302, 2008.

\bibitem{supeene2008deformation}
G.~Supeene, C.~R. Koch, and S.~Bhattacharjee, ``Deformation of a droplet in an
  electric field: nonlinear transient response in perfect and leaky dielectric
  media,'' {\em Journal of colloid and interface science}, vol.~318, no.~2,
  pp.~463--476, 2008.

\bibitem{becker1991experimental}
E.~Becker, W.~Hiller, and T.~Kowalewski, ``Experimental and theoretical
  investigation of large-amplitude oscillations of liquid droplets,'' {\em
  Journal of Fluid Mechanics}, vol.~231, pp.~189--210, 1991.

\bibitem{shiryaeva2003internal}
S.~Shiryaeva, ``On internal mode resonance in a nonlinearly vibrating
  volumetrically charged dielectric drop,'' {\em Technical Physics}, vol.~48,
  no.~2, pp.~152--164, 2003.

\bibitem{hinds1999aerosol}
W.~C. Hinds, {\em Aerosol technology: properties, behavior, and measurement of
  airborne particles, Chapter 3}.
\newblock John Wiley \& Sons, 1999.

\bibitem{ji2010characterization}
X.~Ji, O.~Le~Bihan, O.~Ramalho, C.~Mandin, B.~D’Anna, L.~Martinon,
  M.~Nicolas, D.~Bard, and J.-C. Pairon, ``Characterization of particles
  emitted by incense burning in an experimental house,'' {\em Indoor Air},
  vol.~20, no.~2, pp.~147--158, 2010.

\bibitem{shoyama2018particle}
M.~Shoyama, T.~Kawata, M.~Yasuda, and S.~Matsusaka, ``Particle electrification
  and levitation in a continuous particle feed and dispersion system with
  vibration and external electric fields,'' {\em Advanced Powder Technology},
  vol.~29, no.~9, pp.~1960--1967, 2018.

\bibitem{zuo2016particle}
Z.~Zuo, J.~Wang, Y.~Huo, H.~Liu, and R.~Xu, ``Particle motion induced by
  electrostatic force of a charged droplet,'' {\em Environmental Engineering
  Science}, vol.~33, no.~9, pp.~650--658, 2016.

\bibitem{cho1964contact}
A.~Cho, ``Contact charging of micron-sized particles in intense electric
  fields,'' {\em Journal of Applied Physics}, vol.~35, no.~9, pp.~2561--2564,
  1964.

\bibitem{matsusaka2010triboelectric}
S.~Matsusaka, H.~Maruyama, T.~Matsuyama, and M.~Ghadiri, ``Triboelectric
  charging of powders: A review,'' {\em Chemical Engineering Science}, vol.~65,
  no.~22, pp.~5781--5807, 2010.

\bibitem{shoyama2019mechanism}
M.~Shoyama and S.~Matsusaka, ``Mechanism of disintegration of charged
  agglomerates in non-uniform electric field,'' {\em Chemical Engineering
  Science}, vol.~198, pp.~155--164, 2019.

\bibitem{morsi1972investigation}
S.~Morsi and A.~Alexander, ``An investigation of particle trajectories in
  two-phase flow systems,'' {\em Journal of Fluid mechanics}, vol.~55, no.~2,
  pp.~193--208, 1972.

\bibitem{adamiak2001deposition}
K.~Adamiak, A.~Jaworek, and A.~Krupa, ``Deposition efficiency of dust particles
  on a single, falling and charged water droplet,'' {\em IEEE Transactions on
  Industry Applications}, vol.~37, no.~3, pp.~743--750, 2001.

\bibitem{Srinivasula2022airborne}
P.~Srinivasula and R.~Thaokar, ``Numerical study of airborne particle dynamics
  in vortices subject to electric field,'' {\em physics.flu-dyn}, 2022.

\bibitem{davenport1978field}
H.~M. Davenport and L.~K. Peters, ``Field studies of atmospheric particulate
  concentration changes during precipitation,'' {\em Atmospheric Environment
  (1967)}, vol.~12, no.~5, pp.~997--1008, 1978.

\bibitem{lin2010efficient}
G.-Y. Lin, C.-J. Tsai, S.-C. Chen, T.-M. Chen, and S.-N. Li, ``An efficient
  single-stage wet electrostatic precipitator for fine and nanosized particle
  control,'' {\em Aerosol Science and Technology}, vol.~44, no.~1, pp.~38--45,
  2010.

\bibitem{phadke2021novel}
K.~S. Phadke, D.~G. Madival, J.~Venkataraman, D.~Kundu, K.~Ramanujan, N.~Holla,
  J.~Arakeri, G.~Tomar, S.~Datta, and A.~Ghatak, ``Novel non intrusive
  continuous use zebox technology to trap and kill airborne microbes,'' {\em
  Scientific reports}, vol.~11, no.~1, pp.~1--9, 2021.

\bibitem{bayless2004membrane}
D.~J. Bayless, M.~K. Alam, R.~Radcliff, and J.~Caine, ``Membrane-based wet
  electrostatic precipitation,'' {\em Fuel processing technology}, vol.~85,
  no.~6-7, pp.~781--798, 2004.

\bibitem{tepper2007electrospray}
G.~Tepper, R.~Kessick, and D.~Pestov, ``An electrospray-based, ozone-free air
  purification technology,'' {\em Journal of Applied Physics}, vol.~102,
  no.~11, p.~113305, 2007.

\bibitem{Pramodt2022patent}
P.~Srinivasula and R.~Thaokar, ``Electrohydrodynamic self cleaning air
  purification system,'' {\em Indian Patents application No. XX}, no.~0, p.~0,
  2022.

\bibitem{go2007ionic}
D.~B. Go, S.~V. Garimella, T.~S. Fisher, and R.~K. Mongia, ``Ionic winds for
  locally enhanced cooling,'' {\em Journal of Applied Physics}, vol.~102,
  no.~5, p.~053302, 2007.

\bibitem{go2008enhancement}
D.~B. Go, R.~A. Maturana, T.~S. Fisher, and S.~V. Garimella, ``Enhancement of
  external forced convection by ionic wind,'' {\em International Journal of
  Heat and Mass Transfer}, vol.~51, no.~25-26, pp.~6047--6053, 2008.

\bibitem{jiang2022numerical}
Z.~Jiang, Y.~Gan, and Y.~Shi, ``Numerical analysis on the heat/mass transfer to
  a deformed droplet under a steady electric field,'' {\em International
  Journal of Heat and Mass Transfer}, vol.~188, p.~122617, 2022.

\bibitem{carleson1983effect}
T.~E. Carleson and J.~C. Berg, ``The effect of electric fields on the
  absorption of pure sulfur dioxide by water drops,'' {\em Chemical Engineering
  Science}, vol.~38, no.~6, pp.~871--876, 1983.

\bibitem{phasefCOMSOL}
COMSOL, ``Theory for the three-phase flow interface,'' {\em COMSOL
  Multiphysics® v. 5.0.}, pp.~301--307, 2014.

\bibitem{tomar2007two}
G.~Tomar, D.~Gerlach, G.~Biswas, N.~Alleborn, A.~Sharma, F.~Durst, S.~W. Welch,
  and A.~Delgado, ``Two-phase electrohydrodynamic simulations using a
  volume-of-fluid approach,'' {\em Journal of Computational Physics}, vol.~227,
  no.~2, pp.~1267--1285, 2007.

\end{thebibliography}
\end{document}